\documentstyle[prl,aps,epsfig]{revtex}
\begin{document}
\title{ Particle Annihilation in Cold Dark Matter Micropancakes}
\author{Craig J. Hogan}
\address{Astronomy and Physics Departments, 
University of Washington,
Seattle, Washington 98195-1580}
\maketitle
\begin{abstract}
Cold primordial particle dark matter forms with a distribution in six-dimensional
phase space closely approximating  a three-dimensional sheet. Folds in the 
mapping of this sheet onto configuration space create  ubiquitous sheetlike caustics  
(``micropancakes'').  A typical WIMP dark matter halo has many micropancakes,  
each with a scale comparable to the halo itself,  a width about $10^{-8}$ of the halo 
size and a typical  maximum density up to  about $10^4$ times the halo mean.   
It is suggested here that the total annihilation rate of dark matter particles is dominated by
particles close to these micropancakes, so radiation is emitted predominantly from highly contorted
two-dimensional surfaces rather than a filled volume. The total annihilation rate of particles is
about a factor of 5 higher than predicted from N-body simulations, which cannot resolve
these features. Micropancakes also produce  sharp line discontinuities in the surface brightness of
annihilation radiation.
\end{abstract}
\section{ Annihilating Cold Dark Matter}
Many candidates for particle dark matter separate out of the early 
expansion symmetrically, with 
equal numbers of particles and antiparticles that slowly annihilate at late times. 
For some of the most plausible     weakly interacting massive   particle 
(``WIMP'' \cite{leeweinberg,jungman96,berg1}) candidates,  such as neutralinos,
these   annihilations produce GeV to TeV 
photons and other observable  energetic species at a rate which is potentially 
observable and can be used to constrain the particle parameters
\cite{gunn78,stecker78,silkS84,turner86,silkB87,bouquet89,lake90,berg2,baltz99,gondolo99a,baltz00,calcaneo00}.
Annihilations may be a source of high energy photons already seen in  {\it  EGRET} data,
 including diffuse
emission (e.g. Ref.~\cite{dixon98}),  as well as emission from    the
center of the Galaxy ~\cite{Mayer98}  and  unresolved ``discrete sources''
~\cite{hartman}. Gamma ray backgrounds will be studied with 
better sensitivity and resolution
with future experiments such as {\it GLAST} ~\cite{glast} and {\it VERITAS} ~\cite{veritas}.

The  physics of the annihilation is straightforward
to calculate within the framework of any given particle candidate \cite{jungman96,berg1}, but
because the annihilation rate is proportional to the square of the particle density $n$, the 
calculated rate of annihilations and their spatial distribution depend on
astrophysically complex features of 
the detailed small-scale  spatial distribution of the dark matter particles. The best
calculations\cite{calcaneo00} use high-resolution N-body simulations to compute the dark matter
distribution including   clumpy relic substructure within halos\cite{ghigna98,mooretexas} and   high
density central cusps\cite{ghigna98,mooretexas,carlberg94,navarro96,moore98a}, both of which are
important to the annihilation rate and the expected appearance of the radiation on the sky. This
paper suggests that
    even smaller  predicted structures, sheetlike caustics or ``micropancakes'',
which are not included in even the best   N-body calculations,   significantly increase 
the total annihilation rate and change  the predicted   fine-scale spatial
distribution of emission.

\section{Micropancakes of Cold Dark Matter}
  CDM particles are created in the early universe as a  smooth Hubble flow
 ($\vec v=H\vec
r$)   with   small random velocities,  and therefore occupy a  
  sheet in phase space that is  thin  in three of
the six phase space dimensions.    Liouville's theorem
guarantees that during subsequent collisionless evolution,  the  fine grained  distribution   
 always remains one continuous unbroken thin
sheet, although it  wraps up in a complicated pattern once nonlinear structures form. 
In the emptiest parts of intergalactic space today, the sheet still passes
through each point in space only once; in collapsed systems, it passes many times, so that
the velocity distribution at any point in a CDM halo better approximates a 
discrete rather than a continuous function \cite{sikivie1}.

The wrapping of the phase space sheet (see figures 1 and 2) creates   singularities 
  in the mapping from phase space
onto configuration space \cite{hoganlens,tremaine}.
The most generic are the two-dimensional sheetlike ``fold catastrophes'', although there are also 
  higher order catastrophes at their intersections.  We will 
call these
   two-dimensional caustics    ``micropancakes'' (in contrast to
the much larger Zeldovich pancakes or ``{\it blini}''\cite{zeldovich} which  form in hot-dark-matter
cosmologies). They occur at the surfaces  in space where the number of 
phase sheets changes.

 The density $n$ of particles near a micropancake follows the universal profile of a fold
catastrophe;  it formally diverges  on one side according to the universal scaling
$n\propto
\delta\ell^{-1/2}$ where $\delta\ell$ is the normal distance to the micropancake,
and drops discontinuously on the other side as the number of projected
sheets drops by two.
The physical width of the  density discontinuity is  
 limited only by the fine-grained width of
the  phase sheet--- which is in turn fixed by the 
primordial particle velocity dispersion of the dark matter.

Physically,  micropancakes  form
at the turning points of the particle flow in the frame moving locally with the particles.
They appear in exactly symmetric (spherical, cylindrical or planar) solutions of  halos and
voids
(e. g. \cite{fillmoregoldreich1,fillmoregoldreich2,bertschinger}),
but  
  micropancakes are not an artifact of  any symmetry; they  are  a robust and
generic prediction  of the CDM scenario.
An isolated spherically symmetric halo forms new
micropancakes like tree rings, one per ``year'' (or gravitational crossing time); for typical CDM 
halos with  a rich substructure, the older substructures themselves also contain a  highly
wrapped   structure with hundreds of folds.

 Although   freshly formed micropancakes are   
associated with the sheets and filaments defining the ``cosmic web''\cite{cosmicweb},
they are not the same thing; the micropancakes are pathological sites involving only a 
vanishing portion of the matter at very high density at any given time, whereas the cosmic web
structures define the organization of a large fraction of the material. 
 A
 fairly accurate  visual metaphor is the
pattern of sunlight in a swimming pool: most of the light is cast into a weblike
pattern, but only a small fraction into bright caustic lines.

Indeed, the main reason  why this rich structure has received rather little
attention in the literature is that it has few observable effects 
on astrophysical phenomenology. The micropancakes contain only
a small amount of mass at high density, and have relatively little  effect on the smooth
gravitational potential or on the dynamics of observable stars  and gas \cite{moorecaustics}  (but
see
\cite{sikivie}).  For the same reason, they do not appear in N-body simulations.
However,  the preservation of the coherent cold phase space streams which 
necessarily give rise to
the micropancakes at their turning points has been demonstrated  in simulations (for example,
by constructing Poincar\`e projections showing distribution of particles
in isolated islands in phase space \cite{moorecaustics}), so
we know that micropancakes are indeed predicted by the  physical CDM model. In principle, a
detailed study of these could be used to estimate many properties of the 
micropancakes more accurately than the rough estimates presented here.

In spite of their negligible overall dynamical effects, there
are a few phenomena where the  pathological
properties of  micropancakes
may be   observable. 
These phenomena  are of particular interest
  as a probe of the nature   of the dark matter, since 
the finite width of the caustics   preserves information about the primordial fine-grained
phase density.

\section{Micropancakes in Halos}

Many   real  astrophysical  systems display something like 
phase-wrapping folds\cite{tremaine}, but
particle CDM (and in particular   WIMP) micropancakes
are unusually good   approximations to  ideal  
catastrophes over an exceptionally large dynamic range. Because   the particles 
form with a very high microscopic phase density,
 their caustics create  a local physical density enhancement of
many orders of magnitude over the halo mean.

Suppose a candidate of mass $m_X\approx 1$GeV separates from the thermal
cosmological  plasma
at a temperature of $T_{sep}\approx $0.1 GeV (typical for Lee-Weinberg
weakly interacting dark matter). 
The fine-grained primordial phase space density $Q_P\equiv \rho\sigma_X^{-3}$ is
determined by the physics of the dark matter: the momentum
distribution of the dark matter particles is fixed at $T_{sep}$. 
Normalized to the present density of
dark matter $\rho_0\approx 0.3\rho_{crit}$, 
\begin{equation}
Q_P\approx 10^{36}{\rho_0\over c^3}\left({m_X\over {\rm GeV}}\right)^{3/2} 
\left({T_{sep}\over 0.1{\rm GeV}}\right)^{3/2}.
\end{equation}
This phase density is conserved during adiabatic expansion, so   
the  velocity dispersion $\sigma_X$ at a
 redshift $z$, in units with $c=1$, is about
\begin{equation}
\sigma_{X}(z)\approx 10^{-11.5} (m_X/{\rm GeV})^{-1/2} (T_{sep}/0.1{\rm GeV})^{-1/2} [(1+z)/3] 
\end{equation} 
in the absence of any clustering.
For   collisionless particles, the fine-grained 
density of the phase sheet retains the primordial value $Q_P$
even after nonlinear collapse
\cite{hogandalcanton,dalcantonhogan}. For standard WIMPs, the collisionless
approximation is very good; gravitational two-body scattering is negligible, and
other processes such as annihilation and scattering involve weak interactions
that are negligible after $T_{sep}$.

The evolution of the phase sheet in the nonlinear CDM hierarchy is very 
complex; for this discussion, we only estimate the 
properties in smooth halos characterized by a single density $\rho_{halo}$ and velocity 
dispersion $\sigma_{halo}$. We assume that there  are few
hard scattering centers, so that the main mixing of the phase sheet is   coarse-grained
violent relaxation in halos on each scale of the hierarchy.
There is then only one length scale $\ell_{halo}$ in each halo and most of  the orbits
retain much of their coherence (as seen in both simulations and in 
observations of Galactic stellar moving groups\cite{helminature}.)   We assume that the
 large dimensions of the folds
is  mostly  comparable with the scale of   the  halo.
The number of folds $N$
grows only linearly (not exponentially) with time (although a chaotic system
could lead to a more rapid phase space mixing that would tend to erase micropancakes;
see \cite{goodman}).  These
estimates can be  applied to a hierarchy by regarding subhalos as separate bound systems.

  Suppose a 
halo has $N$  folds of the sheet which add up to the total spatial
density $\rho_{halo}$; then from mass conservation, the space density of
each single sheet is about $\rho_{halo}/N$, times factors of the order of unity
depending on the halo profile and geometry. (This description is only 
adequate for real halos over a limited dynamic range in radius. For example, the outermost
regions of a halo still forming out of the expansion would have $N=1$, the density
and radius of the outermost caustic corresponding to the first turnaround
of particle orbits;  both  $N$ and  $\rho_{halo}$   increase at smaller radii.)
Conservation of
local fine-grained phase density  then 
allows us to estimate from this density the  typical fine-grained velocity dispersion
$\sigma_{Xhalo}$ in the wrapped up sheet (far away from the caustics),
\begin{equation}
{\sigma_{Xhalo}\over \sigma_{halo}}
\approx \left({\bar Q\over Q_P}\right)^{1/3}N^{-1/3}
\end{equation}
where $\bar Q\equiv\rho_{halo}/\sigma_{halo}^3$  denotes
the mean coarse-grained phase space density of the halo.

The fine-grained nonlinear
 evolution of the phase sheet can be  described using action-angle 
variables \cite{helmi,helmiwhite}. Although exact results
have only  been derived in symmetric situations, we can again roughly estimate the 
properties of the  caustics in the general case. The finite minimum
spatial width $\ell_{min}$ of the
micropancakes can be estimated by noting  that the turning points of the orbits
for particles in one sheet     have a spatial 
dispersion about the mean location of the sheet 
 proportional to the dispersion in velocity integrated over an
orbital time, hence
\begin{equation}
{\ell_{min}\over \ell_{halo}}\approx {\sigma_{Xhalo}\over \sigma_{halo}}
\approx \left({\bar Q\over Q_P}\right)^{1/3}N^{-1/3}
\approx 10^{-8} N^{-1/3}. 
\end{equation}
Near a micropancake, the velocity dispersion in the normal direction
increases  to a saturated value $\sigma_{sat}$ depending on $Q_P$ and 
the curvature of the sheet in phase space.  For folds curving on the halo scale
of length and velocity, with the two large dimensions both close to the halo size, the
  dispersion   saturates at a value  
\begin{equation}
\sigma_{sat}/\sigma_{halo} \approx  (\ell_{min}/ \ell_{halo})^{1/2}\approx 
(\sigma_{Xhalo}/\sigma_{halo})^{1/2},
\end{equation}
times numerical factors  depending on the geometry of the fold.
The typical   maximum saturated 
density $\rho_{max}$ is then approximately given by $Q_P \sigma_{sat}\sigma_{Xhalo}^2$, or
\begin{equation}
{\rho_{max}\over \rho_{halo}}\approx \left(\ell_{min}\over \ell_{halo}\right)^{-1/2} N^{-1}
\approx \left({Q_P \over \bar Q }\right)^{1/6}N^{-5/6} 
\approx 10^{4} N^{-5/6}.
\end{equation}
The value of $\bar Q$ for real halos varies by over nine orders of magnitude,
from dwarf spheroidals to galaxy clusters 
\cite{dalcantonhogan}; the numerical values here
estimate $\bar Q$ for ``typical'' CDM halos
using $ ({\bar Q/ Q_P} )^{1/3}\approx  (\sigma_{X}(z_{coll}) / \sigma_{halo})$
at a collapse redshift
$z_{coll}\approx 10$, and a velocity  dispersion   $\sigma_{halo}\approx 10^{-3}$.   
   The number of folds is of the order of
 the dynamical age of the halo
  in units of the gravitational crossing time,  $N\approx$
1 to 100 for typical halos. The powers of $N$ in these estimates   change if
we make different assumptions about the micropancake geometry;
such effects can lead to factors of $\approx 2$ difference in the annihilation
rate estimates below.

The density enhancement near the micropancakes within each
halo or subhalo is around a factor of
    $10^4$, although the total contrast is 
less for old halos or dense central regions with many folds. The angular size of the sharp edges is
less than 
$10^{-8}$ of the  halo size, and the dynamic range, the ratio of size to separation,
is about $10^{-8}N^{2/3}\approx 10^{-6}$ (hence the name, ``micropancakes'').

\section{Annihilation rate with Micropancakes}

One possible technique for measuring 
the properties of these tiny caustics   is  ``micopancake microlensing'' ~\cite{hoganlens}. The
projection of micropancakes onto the plane of the sky leads to line discontinuities in surface
density which produce  universal
signatures in the gravitational lensing of  background stars or other sources: 
images of very distant stars occasionally suddenly appear or disappear, and  
extended sources occasionally display mirror-reflected images about a sharp line 
discontinuity. 
Detection of these exotic  effects is  however  well beyond current observational 
capability.

Here we estimate another, possibly more easily   observable consequence  of 
micropancakes:  their influence on dark matter particle annihilation.
Although each particle spends only a small amount of time in micropancakes,
the density is so high that  most of the annihilations take place there. 
The  micropancakes therefore alter the annihilation rate by a large factor. 

Micropancakes  occur on such a small scale that they are not resolved in CDM simulations (this
is even true of the largest, ``freshest'' pancakes, because they occur in those 
relatively underdense parts of
space where CDM codes are designed for speed to have the  poorest resolution).
Because of this, the statistical properties of their spatial distribution are
poorly characterized compared to the other predictions of CDM.   
We know that orbital phase mixing eventually homogenizes phase space; however,
the estimate above suggests that it takes a very large number of orbits
(perhaps as many as $N\approx 10^{4\times6/5}\approx 10^5$ for smooth halos) to
mix thoroughly enough to destroy the micropancakes. 
Until this happens,  micropancakes
are ubiquitous and   a typical particle will participate in  one at a turning point
 about every orbit. This is enough information to estimate
the  magnitude of the effect on annihilation rates.

Suppose  that the probability of annihilation per particle
per time is $n\sigma v$, where $n$ is the number density of 
(anti)particles and $\sigma v$ is some constant determined by the 
physics of the dark matter (e.g., \cite{jungman96}). Normalizing to unit  size, virial velocity and 
orbital time   for the local halo of density $n_0$, close to the pancake the universal fold
catastrophe profile gives $n\approx n_0\delta \ell^{-1/2}$ within distance
$\delta\ell$ of a micropancake. Each particle enters and leaves the pancake
at a turning point of its orbit, so its velocity relative to the frame of
the pancake is $\delta\dot\ell\approx \delta\ell^{1/2}$;   each particle
trajectory in this frame is $\delta\ell(t)\approx t^2$; and the 
integrated annihilation probability is therefore
\begin{equation}
\int dt n \sigma v\approx n_0\sigma v \int dt/t.
\end{equation}
The
integrated annihilation probability is therefore logarithmically divergent as $t\rightarrow 0$ or
$\delta \ell\rightarrow 0$, the moment when the particle passes through 
the  micropancake. Since each particle passes through one
of these caustics every orbit, this means that the overall annihilation integral
is dominated by the particles close to the pancakes. 
An equivalent formulation is to discuss the probability distribution of 
$n$, which has a power-law tail due to matter near micropancakes, $dP(n)/dn\propto n^{-2}$. The 
annihilation rate is then $\int dn n (dP(n)/dn)\propto \int d\ln n$, again displaying
the high-$n$ logarithmic divergence.

This leads   to the
first main conclusion,
\begin{itemize}
\item
The annihilation radiation  comes predominantly from
 contorted two-dimensional surfaces, rather than from a volume-filling source.
\end{itemize}

An important corollary of this is that:
\begin{itemize}
\item
 The mean annihilation rate is  higher
 than 
one would estimate from N-body studies of clumping--- even those of the highest resolution---
by a significant factor.
\end{itemize}
 The exact value of this
factor is not known but  we can estimate roughly   from the
maximum density contrast  of the micropancakes, estimated above. For each virialized halo or
subhalo,  the total rate is higher than one would guess from
the mean local density  by a factor of $\ln(10^4)-(5/6)\ln N\approx 9-(5/6)\ln N$, 
about a factor of 5.  The bulk of the integral comes far below the resolution limit of
simulations, from the small fraction of particles closest to the micropancakes.   
The argument holds for all
systems and subsystems within a halo, 
or for different radii within a halo, although the
$\ln N$ reduction reflects a   smaller enhancement factor in higher
density regions  (as  the 
  density enhancement in pancakes is diluted by wrapping).
In the absence of micropancakes, the bulk of  
annihilations are predicted to occur in a small fraction of mass in the dense 
parts of halos; this argument indicates that within those dense parts,
 the micropancakes dominate.

\section{Line discontinuities in brightness}
If the CDM annihilations produce  direct gamma ray emission
(as opposed to indirect emission via $e^+e^-$),   
the radiating surfaces may be imaged.  For radiation  in a
monochromatic line, a ``velocity cube'' map (two sky directions and a redshift)
would    show  emission concentrated in
 two-dimensional sheets,  giving a direct three-dimensional image of the 
micropancakes. 
However, if the annihilations produce continuum radiation
or if the  velocity resolution is insufficient to separate the sheets, we should
calculate instead the mean surface brightness of the annihilation radiation.  This
is given by the line-of-sight integral
\begin{equation}
I=\int dz n^2(z) \sigma v.
\end{equation}

Consider now the surface brightness near the line in the sky
 where a micropancake lies normal to the plane of the sky, and
is seen edge-on. Choose coordinates which 
place the line at $\ell=0$ in the plane of the sky, and  at $ z=0$ along the 
line of sight. The line-of-sight
distance to the sheet at a  projected distance $\ell_p$ is given
(again, in locally appropriate units) by $z=\ell_p^{1/2}$ ---
 another  
fold catastrophe, this time  of the mapping of the 2D micropancake sheet onto the sky
(rather than the 3D  phase space sheet  onto 3D configuration space).
 The density of material is thus $n=(\ell_p-z^2)^{-1/2}$, and
the projected surface
brightness at
$\ell_p\ge 0$   becomes
\begin{equation}
I\approx 2n_0^2\sigma v\int_0^{\ell_p^{1/2}}{dz\over  \ell_p-z^2}\approx 2n_0^2\sigma
v={\rm const}(\ell_p),
\end{equation}
while of course there is no flux at $\ell_p<0$.
 Thus, the  folded micropancakes create   abrupt
discontinuities in surface brightness:
\begin{itemize}
\item
 The sky projection of direct emission from a halo of annihilating particles is
  laced by sharp line discontinuties in brightness, the number $N$ of them approximately equal
to the dynamical age.
\end{itemize}

For WIMP dark matter, detailed study of the annihilation radiation background
may thus contain information complementary to that obtainable by direct detection
experiments. However, the experimental challenges are daunting.
Verifying the existence of the discontinuities (as opposed to resolving them)
requires enough resolution to separate the projected lines.
For the most part the lines are closer   than $N^{-1}$ times the
 projected separation scale of galaxies at the Hubble length, on the order
of  arcseconds or smaller (although a small fraction of the
light comes from much larger local halos, including some long caustic threads from our own Galaxy
stretching across the sky).  The amplitude  or fractional contrast  of each line discontinuity is
small, of the order of $N^{-1}$;   
and as shown by the
N-body studies,  most of the flux is coming from dynamically older,  compact substructures within
halos (which have large  $N\ge 10^2$). 

 Measuring the width of the discontinuity, as would be necessary
to actually measure $\bar Q$ directly, is even more challenging.
As discussed above,
it is set by the intrinsic width of the CDM phase sheet; for
WIMP dark matter, this corresponds to an angle of about $10^{-8}$ of the angular
size of the halo under
study. Such a small-scale, low contrast angular structure will not be resolved by  
currently planned GeV-to-TeV  background mapping experiments.

\acknowledgements
I am grateful for conversations with R. Blandford, A. Helmi,  M. Kamionkowski,
 B. Moore, T. Quinn, S. White, and the referee.
This work was supported at the University of Washington
by   NSF grant AST-9617036.

\section{references}
{}

\begin{figure}[htbp] 
%\vspace{1.5in}
\centerline{\epsfig{height=6in, file=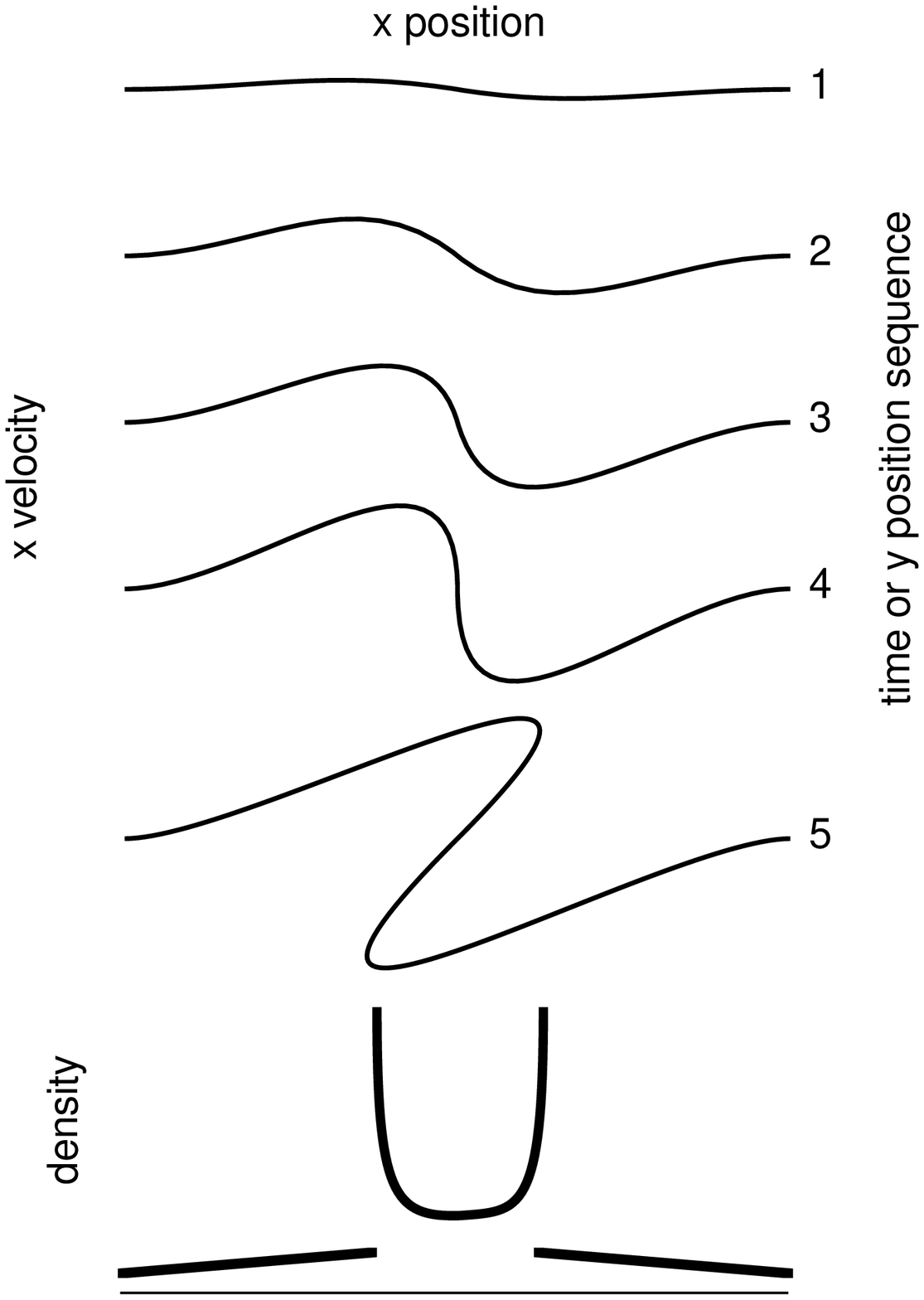}}
\vspace{0.25in}
\caption{ \label{fig: fold} Generic behavior of 
CDM sheet in phase space during formation of a single fold catastrophe. 
Curves show  two dimensions of six-dimensional phase space, 
say $x$ and the $x$ component of velocity, $v_x$. The sequence
of curves can be regarded as a time sequence as a fold forms, or
as a $x,v_x$ cross section at various spatial positions $y$.
The fold caustics formed in step 5  project formally infinite
density in configuration space; 
a density profile cut along $x$ is shown. The actual width and maximum density 
in the ``singularity''  are limited by the
intrinsic  (primordial) width of the CDM phase sheet.   }
\end{figure}

\begin{figure}[htbp] 
%\vspace{1.5in}
\centerline{\epsfig{height=6in, file=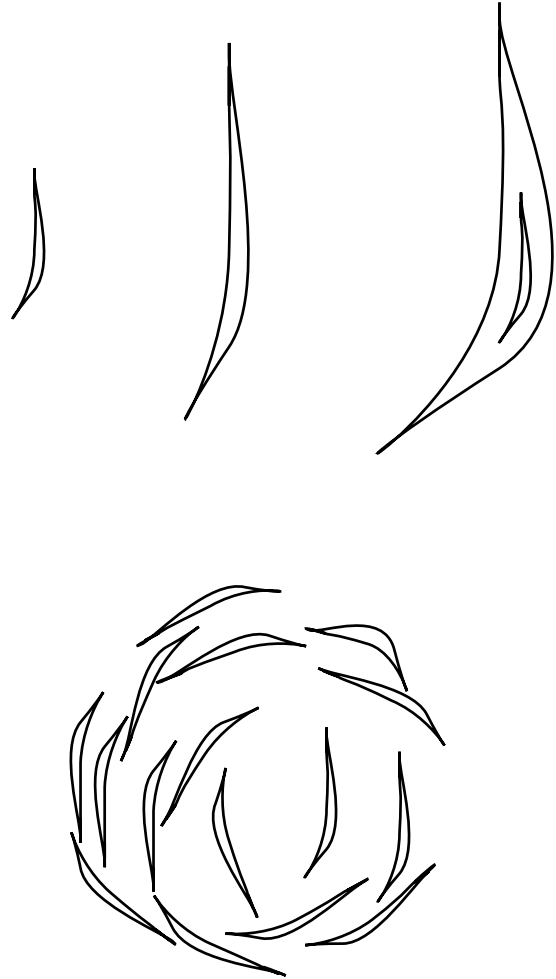}}
\vspace{0.25in}
\caption{ \label{fig: formingfold} Sketch of the formation of micropancakes,
showing the topology of the singularities in configuration space.
The first nonlinear collapse ( with 2D phase space shown in figure 1) forms a small ravioli-like
structure of two micropancakes (fold singularities) joined around the edges (terminating in 
cusp singularities), which then grows. If the structure is bound, 
 a second ravioli forms inside the first (now a calzone) after another
orbit, and so on.
In general a violently-relaxed CDM halo (or subhalo) is threaded with many micropancakes at turning
points of   orbital streams, showing some coherence but with no particular symmetry. This
microstructure appears within each  satellite  of a CDM galaxy halo.}
\end{figure}

\end{document}